\documentclass[aps,
superscriptaddress]{revtex4}

\usepackage[dvips]{graphicx}
\usepackage{amssymb}
\usepackage{amsmath}

\begin{document}

\title{Superconducting Properties of Atomic-Disordered Compound MgCNi$_3$}

\author{A. Karkin}
\email[Corresponding author. E-mail:]{karkin@uraltc.ru}
\author{B. Goshchitskii}
\author{E. Kurmaev}
\affiliation{Institute of Metal Physics, UB RAS, 620219, Ekaterinburg, 
GSP-170, Russia}
\author{Z. A. Ren}
\author{G. C. Che}
\affiliation{National Laboratory for Superconductivity, 
Institute of physics, Chinese academy of Sciences, P. O. Box 603, 
Beijing 100080, P. R. China}
\date{\today}

\begin{abstract}
The effect of radiation-induced disordering in a nuclear reactor 
(fast neutrons fluence $\Phi = 5\cdot10^{19}$ cm$^2$, 
$T_{\text{irr}}$ = 340 K) on resistivity $\rho$, 
superconducting transition temperature $T_C$ and upper critical field 
$H_{C_2}$ of 
polycrystalline MgCNi$_3$ samples was investigated. It was found that $T_C$ 
decreases under irradiation from 6.5 to 2.9 K and completely recovers after 
annealing at 600 $^\circ$C. Temperature dependences $\rho(T)$ are 
characteristic of compounds with strong electron-phonon interaction. 
The $dH_{C_2}/dT$ behaviour 
testifies to a considerable decrease in density of electronic state at 
Fermi level $N(E_F)$ in the course of disordering.
\end{abstract}
\maketitle

Radiation-induced disordering caused by irradiation with 
high-energy particles is a unique method of investigating 
the properties of superconducting and normal states of 
ordered crystals \cite{1,2}. Even in broad-band metals, such 
as intermetallic compounds with A15 structure, long-range 
ordering loss leads to considerable rearrangement of the 
electronic spectrum, resulting in disappearance of individual 
features of the electronic structure. Disordering causes 
decrease in densities at Fermi level $N(E_F)$ and respective 
noticeable drop of $T_C$ in compounds with high initial 
$N(E_F)$ (Nb$_3$Sn or V$_3$Si), and considerable (from 1.5 to 7 K) 
rise of $T_C$ in compounds with low $N(E_F)$ and $T_C$ due to growth 
of $N(E_F)$ (Mo$_3$Si and Mo$_3$Ge) \cite{3,4,5}. In type HTSC compounds, 
disordering leads to more significant changes in properties: 
fast and complete $T_C$ degradation is accompanied with $N(E_F)$ 
decrease and metal-insulator transition \cite{2}. Thus investigation 
of response of a system to radiation-induced disordering serves 
as a kind of a test to reveal the characteristic features of its 
electron states. It was shown in recent papers \cite{6,7} that 
$T_C$ drop from 38 to 5 K observed at MgB$_2$ under radiation-induced 
disordering is connected mainly with considerable drop of $N(E_F)$, 
similar to Nb$_3$Sn or V$_3$Si compounds. In our investigation, we 
concentrated on the effect of disordering on the properties of 
superconducting compound MgCNi$_3$ ($T_C \sim 8$ K) with perovskite cubic 
structure of type SrTiO$_3$, unconventional for intermetallides \cite{8}. 
Our interest in this system was explained by the fact that its 
ground state is close to ferromagnetic due to the presence of a 
narrow peak in $N(E)$ located 45 meV below the Fermi level \cite{9}. 
This allowed us to regard it as a candidate for an unconventional 
(possibly triplet) superconductivity, similar to Sr$_2$RuO$_4$ compound. 
It is known that in Sr$_2$RuO$_4$, as distinct from conventional 
superconducting compounds (intermetallides), $T_C$ undergoes 
anomalously strong suppression even under a slight disorder \cite{10}. 
In MgCNi$_3$, maximum $T_C$ is achieved at excess of carbon content only 
(nominal composition MgC$_{1.5}$Ni$_3$), even though, according to neutron 
diffraction study, the actual composition is closer to Mg$_{0.96}$CNi$_3$, 
and excess carbon occupies the region between sample grain boundaries 
\cite{11}.
	
In the sample preparation, fine powders Mg, C and Ni with purity 
better than 99.5\% were used as starting materials. The mixtures 
of appropriate composition were pressed into pellets; the pellets 
were wrapped in Ta foil and enclosed in an evacuated quartz tube, 
placed in a furnace, heated to 950 $^\circ$C at a rate of 150 $^\circ$C/h and 
kept at this temperature for 5 h, followed by furnace-cooling to 
room temperature. The highest $T_C = 6.5$ K and the best 
superconducting transition corresponded to the nominal 
composition $x = 1.45$ \cite{12}. 
Samples $0.5\times 1\times 5$ mm$^3$ in size were 
irradiated with fast neutrons at $T_{\text{irr}} = (330 \pm 10)$ K, then 
annealed during 20 min at temperatures $T_{\text{ann}}$ from 100 to 
600 $^\circ$C in step of 100 $^\circ$C. 
Resistivity $\rho(T)$ in fields up to 13.6 T was 
measured using a standard four-probe method.
	
\begin{figure}
\includegraphics[keepaspectratio,width=3in,angle=270]{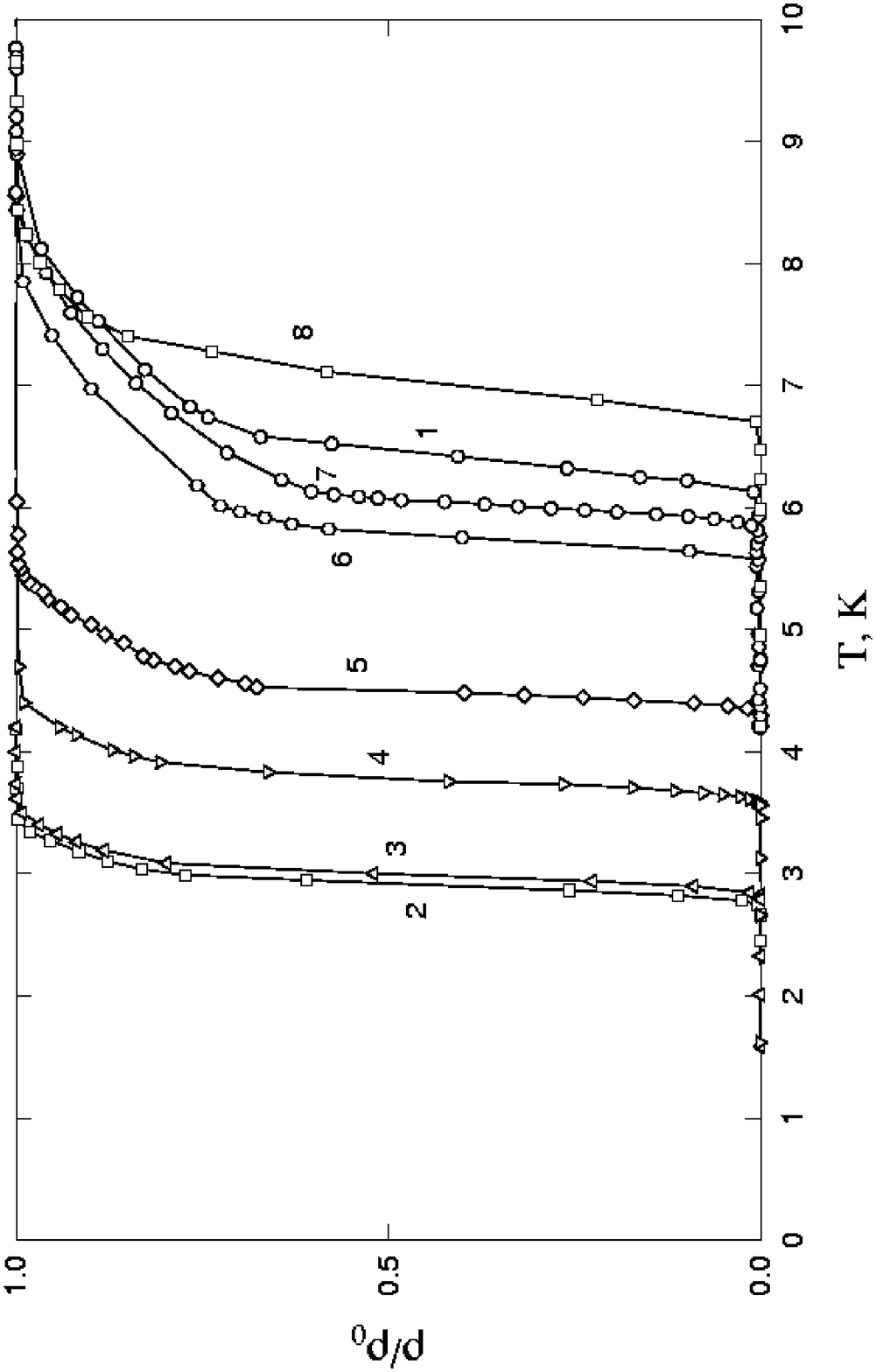}
\caption{Temperature dependences of reduced resistivity $\rho/\rho_0$ 
of initial MgCNi$_3$ sample (1), sample irradiated under fast neutrons fluence 
$\Phi = 5\cdot10^{19}$ cm$^{-2}$ (2) and sample annealed at 
T = (100 - 600) $^\circ$C during 20 min. (3 - 8). 
Solid lines are drawn across experimental points.}
\label{fig:1}
\end{figure}

The initial sample resistivity curve of transition to 
superconducting state (Fig. \ref{fig:1}) is stretched in the direction 
of higher temperatures, onset is about 8 K. 
Mean transition temperature is 6.5 K. 
We defined the superconducting transition temperature $T_C$ as the 
temperature exhibiting half of the normal-state resistivity.
Irradiation leads to $T_C$ drop 
to 2.9 K, and transition becomes narrower. Annealing at 500 $^\circ$C 
almost completely recovers the initial form of dependence $\rho(T)$, 
while after annealing at 600 $^\circ$C, transition becomes more abrupt 
with a higher $T_C$ = 7.1 K compared with the initial sample.
   
\begin{figure}
\includegraphics[keepaspectratio,width=3in]{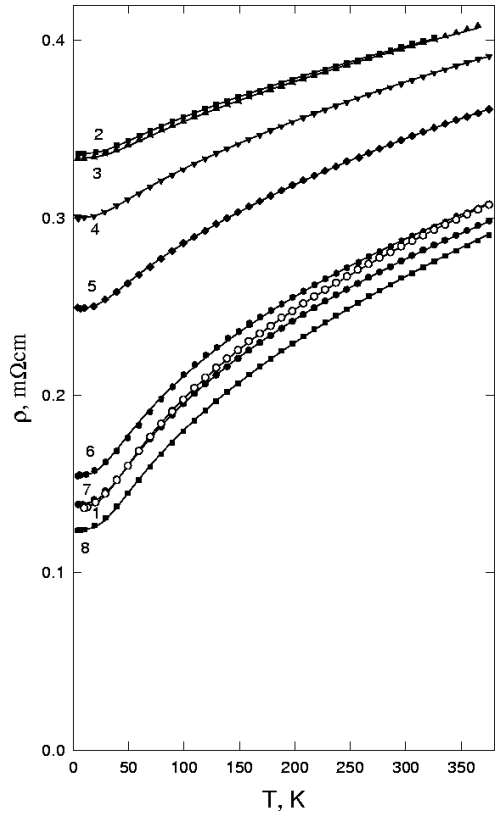}
\caption{Temperature dependences of MgCNi$_3$ sample resistivity 
$\rho(T)$; for designations, see Fig. \ref{fig:1}. Solid lines present the 
calculation using expression (7).}
\label{fig:2}
\end{figure}

Temperature dependences $\rho(T)$ of the initial, irradiated and 
isochronally annealed MgCNi$_3$ samples (Fig. \ref{fig:2}) present curves 
with saturation, typical of the systems with strong electron-phonon 
interaction of types Nb$_3$Sn or V$_3$Si \cite{3}. 
A rather large value of 
residual resistivity $\rho_0 = 0.137$ mOhm$\cdot$ cm (found by 
$\rho$ extrapolation 
to $T = 0$) of a sample in the initial state testifies to an 
insufficient degree of ordering. The absolute value of $\rho(T)$ 
approximately coincides with the data in \cite{13} and is three 
times higher than in [8], even though temperatures dependences $\rho(T)$ 
are practically identical in all cases. Evidently, after irradiation 
and subsequent annealing at 600 $^\circ$C, further ordering and residual 
resistivity drop to $\rho_0 = 0.124$ mOhm$\cdot$cm occur in the sample.
  
\begin{figure}
\includegraphics[keepaspectratio,width=3in,angle=270]{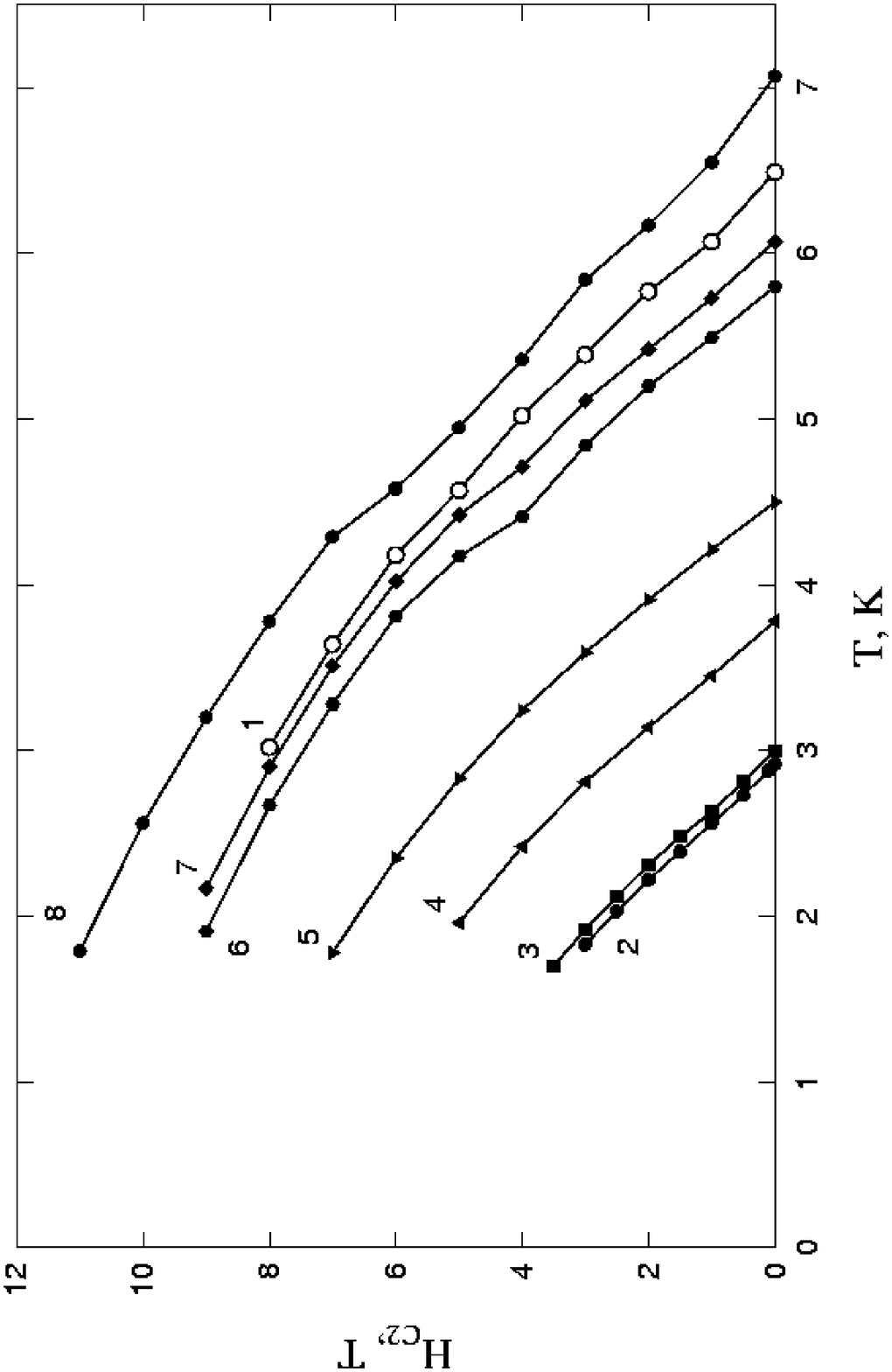}
\caption{Temperature dependences of upper critical field 
$H_{C_2}$ for MgCNi$_3$ sample; for designations, see Fig. \ref{fig:1}. 
Solid lines are drawn across experimental points.}
\label{fig:3}
\end{figure}

The upper critical field $H_{C_2}$, as 
determined from the half-transition temperature (0.5 of the normal-state 
resistivity), has a form typical of second-order 
superconductors (Fig. \ref{fig:3}), the initial sample value of $dH_{C_2}/dT$ 
is in good agreement with the data of paper \cite{13}. 
A relatively 
weak change in the slope of $dH_{C_2}/dT$ should be noted; a very 
similar behaviour at disordering was observed for MgB$_2$ [6]. 
So, for dirty superconductor
\begin{equation}
(-dH_{C_2}/dT)_{\text{dirty}} = (8ek_B/\pi)(1+ \lambda)N(E_F)\rho_0,
\label{1}
\end{equation}
the relatively weak change in $dH_{C_2}/dT$ (Fig. 3) would evidently 
be compensated by a considerable (about 2.5 times) decrease in $N(E_F)$.

Deviations from the Block-Gr\''{u}neisen law
\begin{equation}
\rho(T) = \rho_0 + \lambda_{\text{tr}}F_{\text{BG}}(\theta/T), 							(2)
\label{2}
\end{equation}
defining linear behaviour of $\rho(T)$ at high $T$, where $\theta$ 
is Debye temperature, $\lambda_{\text{tr}}$ is electron-phonon interaction 
constant proportional to parameter $\lambda$ in the McMillan expression
for superconducting transition temperature 
\begin{equation}
T_C \sim  (\omega_{\text{ln}}/1.2)\exp\{- (1+ \lambda)/(\lambda-\mu)\}, 
\quad \mu \sim 0.1,	
\label{3}
\end{equation}
are often described by an empirical expression 
\begin{equation}
1/\rho(T) = 1/\rho_{\text{sat}} + 1/(\rho_0 + \lambda_{\text{tr}}
F_{\text{BG}}(\theta/T)),
\label{4}
\end{equation}
so $\rho(T)$ cannot exceed the value of saturation resistivity 
$\rho_{\text{sat}}$, which for type A15 intermetallides is 
about 0.2 mOhm$\cdot$cm. Intuitive substantiation of (4) boils down 
to the fact that electron scattering becomes inefficient when the 
electron free path $l_{\text{tr}}$ becomes shorter than the Fermi wavelength, 
inversely proportional to wave-vector $k_F$; therefore, in the 
expression for conductivity $\sigma \sim (k_F)^2l_{\text{tr}}$, 
$l_{\text{tr}}$ should be substituted by a value close to $(k_F)^{-1}$. 
The interpolation formula $\sigma\sim (k_F)^2l_{\text{tr}} + k_F$ is 
equivalent to (4).
	
\begin{figure}
\includegraphics[keepaspectratio,width=3in,angle=270]{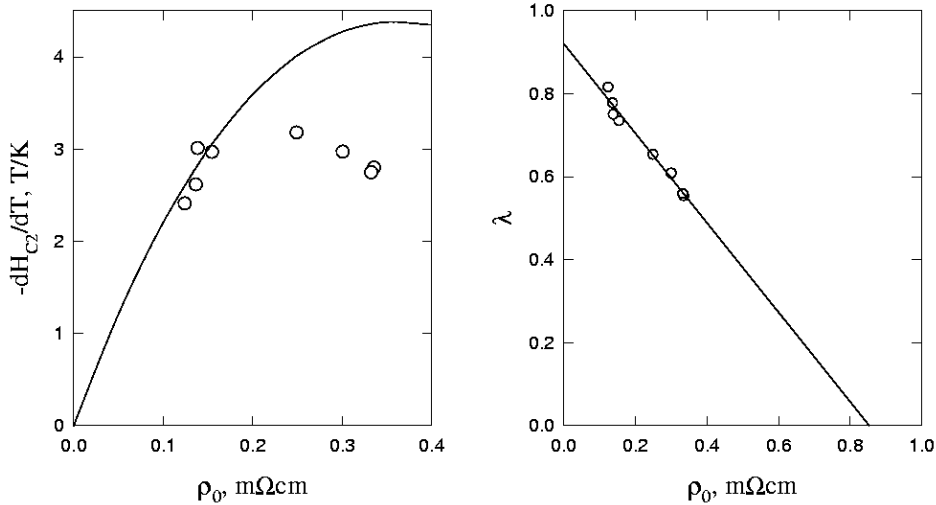}
\caption{Upper critical field derivative 
$-dH_{C_2}/dT$ (left) and 
electron-phonon 
interaction constant $\lambda$ (right) for MgCNi$_3$ sample as a function of 
residual resistivity $\rho_0$. Solid lines present the calculation 
using expressions (8) and (4), respectively.}
\label{fig:4}
\end{figure}

Fitting of experimental data on MgCNi$_3$ to expression (4), 
containing 4 fitting parameters $\rho_{\text{sat}}$, $\rho_0$, 
$\lambda_{\text{tr}}$ and $\theta$, 
yields good agreement with the close values of $\theta = (140 - 155)$ K. 
A similar fitting procedure for MgCNi$_3$ ($T_C \sim$ 8 K) carried 
out in [12] with Einstein, instead of Debye, spectrum, yields the 
following parameters: Einstein temperature $\theta_E = 206$ K, 
$\rho_{\text{sat}} = 0.574$ mOhm$\cdot$cm. 
The obtained value of $\theta$ is noticeably 
lower than that obtained in heat capacity measurements, Debye 
temperature $\theta_D \sim 235$ K [8]. However, using the value 
of $\theta = 150$ K and on the assumption of the Debye spectrum, we obtain 
$\omega_{\text{ln}} = \exp(-1/3)\cdot\theta \sim 105$ K, which is 
considerably lower than 
$\omega_{\text{ln}} \sim 480$ K for MgB$_2$ \cite{14}. 
Expression (3) yields $\lambda \sim 0.8$, 
which compares well with the value of $\lambda\sim 1.1$ for MgB$_2$ [13]. 
Value $\lambda$ as a function of $\rho_0$ (Fig. \ref{fig:4}) may be described 
with a linear dependence
\begin{equation}
\lambda = \lambda_0(1 - (\rho_0/R)),
\label{5}
\end{equation}
where $\lambda_0 = 0.92$, and $R = 0.85$ mOhm$\cdot$cm.

The relatively large value of $\lambda$ (and hence, $\lambda_{\text{tr}}$) 
is generally 
in an agreement with significant nonlinearity of $\rho(T)$ 
characteristic of compounds with strong electron-phonon interaction. 
However, fitting parameter $\rho_{\text{sat}}$, varies significantly from 
0.85 mOhm$\cdot$cm for the initial sample to 0.5 mOhm$\cdot$cm for the 
irradiated sample, which agrees poorly with the meaning of value 
$\rho_{\text{sat}}\sim (kF)^{-1}$, which must be constant in  case of 
a broad-band metal.
	
The origin of $\rho(T)$ "saturation" for systems with strong electron-phonon 
interaction were analyzed in terms of the mean field theory in \cite{15}, 
where it was shown that (in case of a relatively weak coupling which 
does not lead to formation of a pseudogap) scattering rate is 
proportional not to the value of ions r.m.s. displacement 
$\langle u^2\rangle$, 
but rather to $(\langle u^2\rangle)^{0.5}$, and so, in this case, 
instead of (2), we have
\begin{equation}
\rho(T) = \{(\rho_0)^2 + \lambda_{\text{tr}}
F_{\text{BG}}(\theta/T)\}^{0.5},
\label{6}
\end{equation}
which results in type $\rho(T)\sim T^{0.5}$ behaviour at high $T$. 
However, use of (6) fails to yield a satisfactory data description.  
The probable reason is that value $\lambda_{\text{tr}}$, in its turn, 
also depends on 
disordering (is characterized by a sum of static and thermal displacements), 
i.e., on $\rho(T)$; the same reason causes decrease in $\lambda$ with increase 
in $\rho_0$ (Fig. 4). Considering $\lambda_{\text{tr}}$ being in 
dependence on $\rho(T)$, similar to that of $\lambda$ on $\rho_0$ in (4), 
expression (6) is transformed into an equation 
\begin{equation}
\rho(T) = \{(\rho_0)^2 + \lambda_{\text{tr}0}(1 - 
\rho(T)/R_{\text{tr}})(F_{\text{BG}}(\theta/T))\}^{0.5},
\label{7}
\end{equation}
which, when solved for $\rho(T)$, yields the required expression, also 
containing four fitting parameters $R_{\text{tr}}$, $\rho_0$, 
$\lambda_{\text{tr}}$ and $\theta$. 
Expression (7) describes data with the same accuracy as expression (4), 
with similar values of $\theta$, but with almost equal fitting parameters 
$R_{\text{tr}}$ varying within (0.75 - 0.88) mOhm$\cdot$cm. 
Such a good agreement between 
the values of $R$ in (5) and $R_{\text{tr}}$ in (7) does not look casual.

In conclusion, let us consider the probable causes of 
superconductivity degradation in MgCNi$_3$ under disordering. Loss of 
long-range order must lead to smearing of the fine structure of electron 
densities of state; at that, function $N(E)$ smoothes out, but without 
becoming zero. For superconductors with electron-phonon interaction, 
$\lambda \sim N(E_F)$, therefore $T_C$ should never go down exactly 
to zero; the 
latter requirement is evidently satisfied for the majority of compounds 
which may be related to broad-band intermetallides. A qualitatively 
different behaviour is observed in HTSC compounds: in all cases 
superconductivity is completely depressed at a mush higher rate than 
in intermetallides, probably due to non electron-phonon mechanisms of 
superconductivity as well as to a proximity to metal-insulator 
transition \cite{16}.

Value $\lambda$ calculated by expression (3) decreases 1.5 times at 
MgCNi$_3$ under 
irradiation (Fig. \ref{fig:4}), while the above value of $N(E_F)$ 
estimated using 
expression (1) decreases almost 2.5 times. Probably, such discrepancy 
in change of $\lambda$ and $(-dHC_2/dT)$, as it was similarly supposed 
for, e.g., 
MgB$_2$ \cite{6}, may be due to the fact that the dirty limit of 
$l_{\text{tr}} \ll \xi$ 
is not reached in the given region. Coherent length $\xi$ may be estimated 
from the relation
\[
\xi^2 = \Phi_0/\{2\pi(-0.69dH_{C_2}/dT)T_C\},
\]
which yields $\xi = 55$ and 75 $\AA$ for the initial and the 
irradiated samples, respectively. Free path $l_{\text{tr}}$ may be 
estimated from an conventional expression used for conductivity
\[
(\rho_0)^{-1} = (3\pi^2)^{-1/3}(e^2/\hbar)n^{2/3}l_{\text{tr}},
\]
which yields $l_{\text{tr}}\sim 20$ $\AA$ for $\rho_0 = 0.137$ mOhm$\cdot$cm 
(initial sample) 
and $l_{\text{tr}}\sim 8$ $\AA$ for $\rho_0 = 0.337$ mOhm$\cdot$cm 
(irradiated sample). 
These relations of $l_{\text{tr}}$
and $\xi$ are definitely closer to the dirty limit. Further, 
expression (1) allows us to estimate $(-dH_{C_2}/dT)$ using the experimental 
values of $\gamma$ and $\rho_0$ or those obtained from band calculations 
$N(E_F)$. 
According to band calculations \cite{9,17,18}, 
$N(E_F)\sim 2.5$ $(\text{eV}\cdot\text{spin}\cdot\text{cell})^{-1} = 
2.8\cdot 10^{47}$ 
$(\text{J}\cdot\text{m}^3)^{-1}$, using $\lambda\sim 0.8, \quad \rho_0\sim 0.1$ 
mOhm$\cdot$cm, 
we obtain $(-dH_{C_2}/dT)\sim 3$ T/K, which is quite commensurate with 
the experimental value $(-dH_{C_2}/dT)\sim 2.5$ T/K. Thus there are probably 
no reasons to doubt the dirty limit applicability in the given case. 
Assuming $\lambda\sim N(E_F)$, using (1) and (5), we obtain the dependence 
\begin{equation}
(-dH_{C_2}/dT)_{\text{dirty}} \sim 
\lambda(1+ \lambda)\rho_0  = \lambda_0\{1 - (\rho_0/R)\}
(1+ \lambda_0\{1 - (\rho_0/R)\})\rho_0,      		   
\end{equation}
shown as a solid line in Fig. \ref{fig:4}. The causes of noticeable 
deviations at $\rho_0 > 0.25$ mOhm$\cdot$cm are unclear, 
it should be noted only 
that very similar changes in $dH_{C_2}/dT$ at radiation-induced disordering 
were observed for MgB$_2$ [14]. Nevertheless, for MgCNi$_3$, the response to 
disordering is similar to that observed for conventional systems 
(intermetallides) with strong electron-phonon interaction.

\begin{acknowledgments}
Work supported by Minpromnauka, Russia (State Contracts 
No. 40.012.1.1.1150, No. 40.012.1.1.1146/Contract No. 15/02), 
Program of government support to leading scientific schools of 
Russia (Project No. 00-15-96581) and Russian Foundation for Fundamental 
Research (Project No. 01-02-16877).
\end{acknowledgments}

\end{document}